\documentstyle[12pt]{article}
\title{MONOPOLE ANNIHILATION AT THE ELECTROWEAK SCALE---NOT!}
\author{Evalyn Gates, Lawrence M. Krauss\thanks{Also Department
of Astronomy. Research supported
in part by the NSF, DOE, and the Texas National Research Laboratory
Commission. Bitnet: Krauss@Yalehep.}\,, and John
Terning\thanks{Research supported by NSERC.}\\ Center for Theoretical
Physics\\
Sloane Physics Laboratory\\
Yale University\\
New Haven, CT 06511}

\date{February 1992}

\begin{document}
\setlength{\baselineskip}{18pt}
\maketitle\maketitle
\begin{picture}(0,0)(0,0)
\put(300,270){YCTP-P3-92}
\end{picture}
\vspace{-24pt}
\begin{abstract}
\setlength{\baselineskip}{18pt}
\normalsize
We examine the issue of monopole annihilation at the electroweak scale
induced by flux tube confinement, concentrating first on the simplest
possibility---one which requires no new physics beyond the standard
model.
Monopoles existing at the time of the electroweak phase transition  may
trigger $W$
condensation which can confine magnetic flux into flux tubes.
However we show on very general grounds, using several independent
estimates, that such a mechanism is impotent. We then present several
general dynamical arguments constraining the possibility of monopole
annihilation through any confining phase near the
electroweak scale.

\end{abstract}

\vfill\eject

The ``monopole problem" has been with us since the advent of Grand
Unified
Theories (GUTs), which allow the formation of these non-singular stable
topological defects when a
semi-simple gauge group is broken to a lower symmetry group that
includes an explicit
$U(1)$ factor.  These objects typically have a mass $m_M\simeq m_X /
\alpha$, where $m_X$ is the mass of the gauge bosons in the
spontaneously broken
GUT theory, and $\alpha$ is the fine structure constant associated with
the
gauge coupling of the theory.

Shortly after it was recognized that monopoles could result as stable
particles in spontaneously broken GUT models \cite{'tHooft}, and also
that they
would
be produced in profusion during the phase transition associated with the
GUT symmetry breaking in the early universe \cite{Kibble}, it was also
recognized
that they posed a potential problem for cosmology.  Comparing
annihilation rates
with the expansion rate of the universe after a GUT transition, it was
shown
\cite{Preskill,Zeldovich} that the monopole to photon ratio would
``freeze out"
at a level of roughly  $10^{-10}$.  Not only would such an initial
 level result in a cosmic
mass density today which is orders of magnitude larger than the present
upper
limit, but direct observational limits on the monopole abundance in our
neighborhood are even more stringent \cite{parker}.

This cosmological problem was one of the main motivations for the
original inflationary scenario\cite{guth}.  However one of the chief
challenges
to the original
inflationary solution of the monopole problem was the necessity of
having a reheating temperature which is high enough to allow
baryogenesis, but
low enough to suppress monopole production.  In addition, recent
work on large scale
structure,
including observed galaxy clustering at large
 scales, large scale velocity flows, and the absence of any observable
anisotropy in the microwave background, has put strong
constraints on such models.

With the recent recognition that even something as exotic
as
baryogenesis may be possible within the context of the standard
electroweak
theory (supplemented by minor additions), it is worth examining the
issue of
whether the monopole problem may be resolved purely through low energy
physics.  A canonical method by which one might hope to achieve
complete annihilation is by confining monopole-antimonopole pairs in
flux tubes, such as might occur if $U(1)_{em}$ were broken during some
period. Proposals along this line, based on introducing
 new physics have been made in the past,  eg.
\cite{LangPi,Lazarides}.  Most recently, the possibility that such a
phase
might briefly occur near the electroweak breaking scale, for multi-Higgs
models, has also been raised \cite{Sher}. By far the simplest
possibility, however, is
that flux tube confinement of monopoles might occur in the standard
model
unsupplemented by any new physics.  We explore this issue in detail
here, and
then go on to examine the general dynamical obstacles facing any model
involving monopole confinement at the electroweak scale.

\vspace{.1 in}

\noindent{\bf  1. Monopole Confinement in the Standard Model:}

It has been known for some time that the electroweak vacuum
in the broken phase is unstable
in the presence of large ($\geq m_W^2/e$) magnetic
fields\cite{Skalozub}.
The instability is due to the coupling between the magnetic field $H$
and the magnetic moment of the massive $W$ gauge bosons.  Due to this
coupling
the effective mass of the $W$ at tree level is
\begin{equation}
        m_{W_eff}^2 = m_W^2 - eH
\end{equation}
where $e=g \sin\theta_W$  (all expressions are given in
Heaviside-Lorentz units for electromagnetism).  This effective mass
squared becomes
negative
for $H_c^{(1)}\geq m_W^2/e$.
The general resolution \cite{Skalozub}
of the instability is the formation of a condensate of
$W$ and $Z$ bosons, which sets up currents that antiscreen the magnetic
field. The vacuum then acts as an anti-type II superconductor, and
the energy is minimized by the formation of a periodic network of
magnetic flux tubes.  As we shall describe
in some length later, Ambj{\o}rn and Olesen have also shown,
at least for the special case $m_H=m_Z$, that if
the magnetic field increases above
$H_c^{(2)} = {{m_W^2}\over {e \cos^2 {\theta}_W}}$, the full
$SU(2)_L \otimes U(1)$  symmetry is restored \cite{AO1990}.
Thus for an external magnetic field
$H_c^{(1)}< H < H_c^{(2)}$, the electroweak vacuum passes through a
transition region where a $W$ condensate exists and the magnetic
field is confined in a periodic network of flux tubes.

It is possible to imagine how such a phase might arise naturally in a
way which
might lead to monopole-antimonopole annihilation at the electroweak
scale
in the early universe (This idea has also been suggested elsewhere in
the
literature \cite{Owen}).   First of all,
 the magnetic field
necessary to produce such a phase could come from the monopoles
themselves, provided
the electroweak transition is second order.
In this case, the mass of the
$W$ boson generically has a temperature dependence of the form
\begin{equation}
m_W^2 (T) \approx m_W^2(0)[1 - T^2/T^2_c]
\label{m_W}
\end{equation}
where $T_c$ ($\approx 300$ GeV) is the critical temperature associated
with electroweak breaking.
Thus, just below the transition temperature $T_c$, relatively small
magnetic
fields could trigger $W$ condensation.  A remnant density of GUT-scale
monopoles
could provide such
a magnetic field.  Once the condensate forms,
monopoles would become confined to the network of flux tubes, whose
width is
related to the $W$ mass, as we shall describe.
Once the width of the flux tubes is of the same order as the distance
between
monopoles, the monopoles would experience a linear potential and
begin to move towards each other.  If the flux tubes exist for a
sufficiently
long time, the monopoles could annihilate, and their density would
correspondingly decrease.

This picture is very attractive in principle.  However, we now
demonstrate, using a series of arguments which probe this
scenario in successively greater detail, that the
parameters associated with such a transition at the electroweak scale
generally preclude
it from being operational. Moreover, we present dynamical arguments
relevant
for any scenario involving monopole annihilation via flux tube
confinement at the
electroweak scale.

\vspace{.1 in}

\noindent{\bf 2. Kinemetic Arguments: Non-annihilation via Magnetic
Instabilities}:

(a) A Global Argument:  In figure 1, we display a phase diagram
describing the
$W$ condensation picture discussed above, as a function of both
temperature $T$
and
background magnetic field $H$. At $T=0$, for the case examined by
Ambj{\o}rn and Oleson\cite{AO1990,AO1989a,AO1991}, in the region
$1 < He/{m_W^2} <
1/\cos^2\theta_W$ a magnetic flux tube network extremizes the energy and
both the $\phi$ (Higgs) and $|W|^2$ fields develop non-zero
expectation
values.  For finite
temperature the phase boundaries evolve as shown, in response to the
reduction in the
$W$ mass with temperature, up to $T=T_c$, where they meet.
Thus, the phase in
which flux tubes and a $W$ condensate are energetically
preferred falls in between
these two curves.

While the actual magnetic field due to the presence of a density of
monopoles and
anti-monopoles will be complicated and inhomogenous, we first
approximate it by a
homogenous mean field $H_m$, whose precise value is not important for
this discussion.
(As we will later show, given the remnant density of monopoles predicted
to result from a
$GUT$ transition, the value of this field will be well below the zero
temperature
critical field $m_W^2 (0)/e$ at the time of the electroweak phase
transition.)  As the
universe cools from above $T_c$, this background magnetic field will
eventually cross the
upper critical curve for the existence of a flux tube phase.

\begin{sloppypar}
We now imagine that immediately after this happens, flux tubes form, and
monopole
annihilation instantaneously begins. We shall later show that this is
far from the actual
case.  Nevertheless, this assumption allows us to examine constraints on
monopole annihilation
even in the most optimistic case.   As monopole-antimonopole
an\-ni\-hi\-la\-tion pro\-ceeds, the
mean back\-ground mag\-ne\-tic field falls quickly.  At a certain point
this
mean field will fall
{\it below} the lower critical curve, and if it is this background field
which governs the
energetics of $W$ condensation, the $W$
condensate will then become unstable, the
magnetic field lines will once again spread out, and
monopole-antimonopole annihilation will cease.
As can be seen in the figure, the net reduction in the magnetic field
expected from this period of
annihilation will be minimal. Quantitatively the final field (neglecting
dilution due to expansion
during this period) will be a factor of $\cos^2{\theta}_W$ smaller than
the
initial field.  This is hardly
sufficient to reduce the initial abundance of monopoles by the many
orders of magnitude required to
be consistent with current observations.
\end{sloppypar}

(b) A Local Argument:  The above argument points out the central problem
for a
monopole annihilation scenario based on magnetic field instabilities at
the electroweak
scale.  In order to arrange for flux tubes to form, and confinement of
monopoles to occur,
the field must be tuned to lie in a relatively narrow region of
parameter space.
Nevertheless, a potential problem with the above argument, even if it
were less sketchy,
is
that flux tube formation, and monopole annihilation, may more likely be
related to local and not
global field strengths. For example, even if the globally averaged
magnetic
field is
reduced by annihilation, the local field between a monopole-antimonopole
pair connected
by a flux tube may remain above the critical field, so that the
flux tube will presumably persist, and annihilation can proceed.  We now
demonstrate that
even under the most optimistic assumptions about the magnitude of local
fields, for
almost all of electroweak parameter space, local flux tube
formation at a level capable of producing a confining potential between
monopole-
antimonopole pairs will not occur. We first consider the case for which
solutions
(involving a periodic flux tube network) were explicitly obtained by
Ambj{\o}rn and Oleson\cite{AO1989a}.

The area A of flux tubes forming due to the $W$ condensate can be
obtained by
minimizing the classical field energy averaged over each cell in the
periodic
network in the presence of a background $H$ field \cite{AO1989a}:
\begin{equation}
\overline{{\cal E}}_{min}\rm \mit A\rm ={{\mit m}_{W}^{\rm 2} \over \mit
e}
\int_{ce\ell \ell }^{}{f}_{\rm 12}{\mit d}^{\rm 2}\mit x\rm -{{\mit
m}_{W}^{\rm 4} \over
2{\mit e}^{\rm 2}}\mit A\rm +\left({\mit \lambda \rm -{{\mit g}^{\rm 2}
\over 8co{s}^{2}\mit
\theta_W }}\right)\rm \int_{}^{}\left({{\mit \phi }^{\rm 2}-{\mit \phi
}_{\rm
0}^{2}}\right)^2{\mit d}^{\rm 2}\mit x,
\end{equation}
where $f_{12}$ is the magnetic field, and $\lambda$ is the
$\phi^4$-coupling
in the Lagrangian, and $\phi_0$ is the Higgs VEV.
Utilizing the topological restrictions on the flux contained in the flux
tubes (containing
minimal flux $2 \pi/e$),
\begin{equation}
\int_{ce\ell \ell }^{}{f}_{\rm 12}{\mit d}^{\rm 2}\mit x\rm =\oint_{}^
{}\vec{\mit A}\rm \cdot \vec{\mit d\ell }\rm =2\pi \rm /e,
\label{E}
\end{equation}
this yields an expression for A, determined by the energy density
$\overline{{\cal E}}_{min}$, which is in turn a function of the external
magnetic field:
\begin{equation}
 A\rm ={2\pi \rm {\mit m}_{W}^{\rm 2} \over
{\mit e}^{\rm 2}\left[{\overline{\mit \cal E}_{min}\rm +{\mit
m}_{W}^{\rm
4}/2{\mit
e}^{\rm
2}-\left({\mit \lambda \rm -{{\mit g}^{\rm 2} \over 8co{s}^{2}\mit
\theta_W }}\right)\rm
\int_{}^{}\left({{\mit \phi }^{\rm 2}-{\mit \phi }_{\rm
0}^{2}}\right)^2{\mit d}^{\rm 2}\mit
x}\right]}.
\label{A}
\end{equation}
Taking the Bogomol'nyi limit\cite{Bogo}
${\mit  \lambda \rm ={{\mit g}^{\rm 2} \over 8 {\cos}^{2}\mit
\theta_W }}$, corresponding to $m_H=m_Z$, the classical field equations
simplify, and
the properties of the flux tubes can be derived.  In
particular, one can show
\cite{AO1990} that the area of the flux tubes is restricted to lie in
the range
\begin{equation} 2 \pi \cos^2 \theta_W < A m_W^2 <2 \pi .
\label{Aineq}
\end{equation}
{}From
our point of view, it is important to realize that this result is
equivalent to the statement that a $W$ condensate can only
exist between the two
critical values of the magnetic field
\begin{equation}
 {{\mit m}_{W}^{\rm
2} \over \mit e\rm
\cos^2 \mit \theta_W } > {\mit H}^{}\rm > {{m}_{W}^{\rm 2} \over \mit
e}\rm.
\label{H}
\end{equation}
Moreover, it gives a one to one correspondence
between the area of the
flux tube and the background magnetic field value in this range.  We
shall use this
correspondence, both in the Bogomol'nyi limit and beyond, to examine the
confinement properties of
such a flux tube network connecting monopole-antimonopole pairs.

\begin{sloppypar}
Magnetic monopoles are formed at the $GUT$ transition with a density
of about one monopole per horizon volume.  This corresponds to a value
of ${n_M\over s}=10.4 {g_{*}}^{1/2} (T_{GUT}/M_{Pl})^3 \sim
{10^{2}}(T_{GUT}/M_{Pl})^3$,
where $n_M$ is the num\-ber
den\-si\-ty of mo\-no\-poles, $g_{*}$ is approximately the number of
helicity
states in the
radiation at the time $t_{GUT}$, $M_{Pl}$ is the Planck mass,
and $s$ is the entropy of the universe
at this time.
Since $T_{GUT}$ could easily exceed $10^{15}$ GeV for SUSY GUTs, it is
quite possible
that the initial monopole abundance left over from a GUT transition is
${n_M\over s}
>10^{-10}$. Preskill has shown that in this case monopoles will
annihilate shortly after
the GUT transition until ${n_M\over s} \sim 10^{-10}$\cite{Preskill},
and this value
remains constant down to the electroweak scale. Since
$s=(2\pi^2/45)g_{*} T^3$, the
mo\-no\-pole
 num\-ber den\-si\-ty at the elec\-tro\-weak transition ($T_c\sim 300
GeV$) of
$\approx$ 0.13 $GeV^3$ (assuming
$g_{*}(T_{c}) \approx 100$) corresponds to a mean intermonopole
spacing of $L \approx$ 2
$GeV^{-1}$.  From this, we can calculate the mean magnetic field
produced
by the monopoles with Dirac
charge $h = 2\pi/e$. In general, because the monopole background is best
described as a ``plasma"
involving both monopoles and antimonopoles, the mean magnetic field will
be screened at
distances large compared to the intermonopole spacing. However,
because we
will demonstrate that even under the most optimistic assumptions,
monopole-antimonopole annihilation
will not in general occur, we ignore this mean field long-range
screening, and consider the local
field in the region between a monopole-antimonopole pair to be
predominantly that of nearest
neighbors, i.e. a magnetic dipole.  While the field is not uniform in
the region between the monopole
and antimonopole, we will be interested in the minimum value of the
field here.  We shall make the
(optimistic) assumption that if this field everywhere exceeds the
critical value $m_{W}^2/e$ on the
line joining the two monopoles, that an instability of the type
described above, involving a
condensate of $W$ fields and an associated magnetic flux tube,
can occur along this line.
\end{sloppypar}

    For a monopole-antimonopole pair separated by a
distance $L$, the minimum field will be halfway between them, and will
have a magnitude $H=2h/{\pi L^2} =4/{eL^2}$.  For this field to exceed
the
minimum
Ambj{\o}rn-Oleson field $m_{W}^2/e$ then implies the relation:
$L<2/m_{W}$.
For a
value $m_{W}=81 GeV$ this relation is manifestly not satisfied for the
value of $L$
determined above. However, assuming a second order transition, as we
have described, the $W$
mass increases continuously from zero as the temperature decreases below
the critical
temperature, implying some finite temperature range over which the
(fixed) background field
due to monopoles will lie in the critical range for flux tube
formation.  In
this case, the magnetic field would enter this range from {\it
above}.  In order that the magnetic field
lie in the range given by
inequality (\ref{H}), we find
 \begin{equation} 2/m_{W} < L<2
\cos\theta_W/m_{W}
\label{L1}
\end{equation}

Nevertheless, even if a
flux tube forms connecting the monopole-antimonopole pair, this will not
result in a
confining linear potential until the width of the flux tube $2r <
L$.  A bound on
this width
can be obtained from the lower bound on the area of the flux tube
(equation (\ref{Aineq})):

\begin{equation} 2r > 2\sqrt2 \cos\theta_W/m_{W} ~.
\label{width}
\end{equation} when the
magnetic field is at
its upper critical value of $m_W^2/e\cos^2\theta_W$. This implies the
constraint
\begin{equation} L>  2\sqrt 2 \cos\theta_W/m_{W} ~.
\label{L2}
\end{equation}

As can be seen, inequalities (\ref{L1}) and (\ref{L2}) are mutually
inconsistent.  Hence, there  appears to be
no region in which both a Ambj{\o}rn-Oleson type superconducting phase
results, and
at the same time monopole-antimonopole pairs experience a confining
potential.  We expect
the situation will be similar to the quark-hadron phase transition
when the
transition is second order.  In that case, it is impossible to
distinguish between a
dense plasma of confined quarks and a gas of free quarks, because the
mean interquark
spacing is small compared to the confinement scale.  Here there will be
no physical impact
of a short superconducting phase, because the confinement scale is
larger
than the distance
between monopoles required to trigger the phase transition.  We expect
no significant
monopole annihilation during the short time in which this phase is
dynamically favored as
the $W$ mass increases.

This result has been derived in the Bogomol'nyi limit, when $m_H=m_Z$.
What about going
beyond this limit? First, note that the energy density of the external
magnetic
field, ${\cal E} = H^2/2$, provides an upper bound on $
\overline{\mit \cal E}_{min}$. Then from equation (\ref{A})
one can show that as long as  $\lambda >  g^2/8\cos^2\theta_W$
($m_H>m_Z$), the flux tube area, for a fixed value of the field, is {\it
larger} than it is in the Bogomol'nyi limit.   While we have no
analytic estimate of the upper critical field, and hence no lower bound
on the flux tube
area, the scaling between area and magnetic field will still be such
that for a given
monopole-antimonopole spacing, and hence a given magnetic field
strength,
the area of the
corresponding flux tube will be larger than in the Bogomol'nyi limit.
Hence the
inconsistency derived above will be exacerbated.  Only in the narrow
range $m_Z/2\stackrel{<}{\sim} m_H<m_Z$
(still allowed by experiment) is there a remote possibility that even in
principle, flux tube
areas may be reduced sufficently so that confining potentials may be
experienced by monopoles
triggering a $W$ condensate.  However, in this range, the energy
(\ref{E}) can be reduced by
increasing $\phi$, so we expect that instabilities arise in
this range which are likely
to make a $W$ condensate unstable in any case.

\vspace{.1 in}

\noindent{\bf 3. Dynamical Arguments Against Annihilation}:

 Even if a confining potential may be
achieved through flux tube formation, there are dynamical reasons to
expect monopole
annihilation will not be complete.  These arguments apply to any
scenario involving a confining phase for monopoles, and suggest that
estimates based on the efficacy of monopole annihilation
may be
overly optimistic. In
the first place,
we can estimate the energy of a
monopole-antimonopole
pair separated by a string of length $L$.  For a long flux tube of
radius
$r$,
considerations of the electromagnetic field energy trapped in the tube
imply a net energy
stored in the flux tube of
\begin{equation}
E= {L \over 2\alpha r^2 }.
\label{kappa}
\end{equation}
  Considering the
case when $L
\approx 2r$, when confinement would first begin, we find the energy
associated with the
string tension is $E = {2 \over  \alpha L } \approx$ 130 $GeV$.  This is
significantly smaller
than the mean thermal energy associated with a transition temperature
$T_c \approx$ 300
$GeV$.  Hence, if the string tension does not vary significantly over
the
period during
which the magnetic field exceeds the critical field, the
string tension exerts
a minor perturbation on the mean thermal motion of monopoles, and hence
will not
dramatically affect their dynamics.  The only way this would not be the
case would be if
the monopole-antimonopole pair moved towards one another at a rate which
could keep
the magnetic
field between them sufficiently large so as to track the increase in the
minimum critical
field as $m_W(T)$ increased to its asymptotic value.  However, this
cannot in general
occur, because thermal velocities are sufficiently large so as to swamp
the motion of the
monopole-antimonopole towards each other. Using the mean thermal
relative
velocity of  monopoles at
$T=T_c$,  we can calculate
how much time, $\delta t$, it would take to traverse a distance equal to
the
initial
 mean distance between monopoles.  Since the
thermal velocity is
$\ll 1$, non-relativistic arguments are sufficient.
We find $\delta t/t
\approx 4.6 \times 10^{-6}$, for $m_M \approx 10^{17} GeV$, and $T_c
\approx$ 300 $GeV$.  During such a
small time interval, $m_W(T)$ remains roughly constant, and hence so
does the string
tension. We find that during the time $\delta t$ the flux tube induced
velocity of the monopole-antimonopole pair
remains a  small
fraction of the mean thermal
velocity, for $m_M >10^{15} GeV$.  Thus, monopoles and
antimonopoles will
not in general move towards one another as $m_W$ increases.  Since
$r(T)$ will not change
significantly between $H_c^{(1)}$ and $H_c^{(2)}$ as $m_W$ increases, if
the mean
inter-monopole spacing
remains roughly constant, monopole annihilation will, on average, not
proceed before the
field drops below its critical value.

What about the more general case  of
a brief superconducting phase which might result if $U(1)_{em}$ is
broken for a
small temperature range around the electroweak scale
\cite{Sher}?  In this  case, the flux tube
area is not driven by the strength of the background magnetic field, and
hence is not tied
to monopole-antimonopole spacing.  Nevertheless, dynamical arguments
suggest that annihilation,
even in this case, may be problematic.  We describe three obstacles
here: (a) as above, the field energy
contained in
the  string
may not be enough to significantly alter the dynamics of a thermal
distribution of monopoles;  (b) even in the event that this energy is
sufficiently large, the time required to dissipate this energy will in
general
exceed the lifetime of the universe at the time of the  $U(1)_{em}$
breaking transition; (c) the time required for monopoles to annihilate
even once
they have dissipated most of the string energy and are confined within a
``bag"
may itself be comparable to the lifetime of the universe at the time of
the
transition.

(a) Consider the energy (\ref{kappa}) stored in the flux tube.  The
radius, $r$,
will depend upon the  magnitude of the
VEV of the field responsible for breaking $U(1)_{em}$.  If this
symmetry breaking
involves a second order transition, then until this field achieves a
certain minimum
value, flux tubes will not be sufficiently thin to produce a confining
potential for
monopoles.  Moreover, even if this VEV quickly achieves its maximum
value, one must
investigate whether or not this field energy is large compared to the
thermal energy at
that time, in order to determine whether the monopoles will be
dynamically driven towards
each other.  As long as $r^{-1} \approx e{\phi_0} \approx eT_c$, where
in this
case $\phi_0$  represents the VEV if the field associated with
$U(1)_{em}$
breaking
and $T_c$ represents the transition temperature, then $E
\gg  T$, so that the
condition of a confining potential is in general satisfied.
Nevertheless, one must also verify that this
inequality is such that the Boltzmann tail of the monopole
distribution with velocities
large enough to be comparable to this binding energy is sufficiently
small (i.e. that
sufficiently few monopoles have thermal motion which is not
significantly affected by the confining potential).
If we assume that such monopoles do not annihilate, then to avoid the
stringent limits on the monopole density today probably requires $E >
O(30)
T$. Determining $L$ by scaling from the initial density, we find that if
$\phi_0 =
\rho  T_c$, then the ratio of the binding energy to the transition
temperature,
$E/T_c \approx 3800 \rho^2$, independent of $T_c$.  This implies a
rather mild constraint on the VEV of the field associated with
$U(1)_{em}$
breaking: $\rho > 0.09$.

(b) Monopole's must dissipate the large energy associated with the
string
field energy if they are to annihilate.  There are two
possible
ways in which this energy can be dissipated: thermal scattering, and the
emission of radiation
\cite{Preskill,Vilenkin}. Utilizing the estimates of energy loss by
radiation given by Vilenkin \cite{Vilenkin} we find that this process
requires $\approx 10^{15}$ times longer to dissipate the string energy
than
the lifetime of the universe at the time of the
transition.\footnote{This
calculation itself is probably an underestimate (unless the monopole
couples to massless or light particles other than the photon), since it
assumes
the  photon is
massless, which it is not in this phase.}  Hence, we concentrate on the
possibility of dissipating the energy by thermal scattering.   We shall
assume
here that $\rho \approx 1$, so that the initial average monopole-
antimonopole
pair energy is $\approx 3800 T$.  The energy loss by collisions with
thermal
particles in the bath is \cite{Vilenkin}
$ {dE}/dt \approx - bT^2 v^2$, where $b = 3 \zeta(3)/(4
\pi^{2})
\sum {(q_i/2)}^2$, and the sum is over all helicity states of charged
particles
in the heat bath.  At $T \approx$ 100 $GeV$, $b \approx 0.7$.  Utilizing
the
relationship between temperature and time in a radiation dominated
Friedmann-Robertson-Walker universe, we then
find
\begin{equation}
\ln {\left( {E_f \over  E_i}\right)} = {{0.03 b M_{Pl}} \over 2 m_M}
\ln {\left({{t_i} \over {t_f}} \right)}~.
\end{equation}
We will take $E_f$ to be the string energy (\ref{kappa}) when $L=2r$,
i.e.
the energy when
the string has become a ``bag".
This implies that the time required to dissipate the initial string
energy is O(50) $t_i$ for $m_M \approx 10^{17} GeV$.
Unless the phase of broken $U(1)_{em}$ lasts
for longer than this time (which does depend sensitively upon the
monopole mass), not all the string energy will be dissipated.  We
have ignored here possible transverse motion of the string.  This energy
must
also be dissipated by friction, which may be dominated by Aharanov-Bohm
type
scattering\cite{alford}.\footnote{We have
been informed that this issue
is
being treated in detail by R. Holman, T. Kibble, and S.-J.
Rey\cite{sj}.}
In any
case,  this is a
rather severe constraint on the temperature range over which the $U(1)_
{em}$ breaking phase must last.

(3)  Once the string energy is dissipated so that the mean distance
between
monopole-antimonopole pairs is of order of the string width, they will
be
confined in a ``bag", and one must estimate the actual time it takes for
the pair
to annihilate in such a ``bag" state. (The monopole ``crust", of
characteristic size
${m_W}^{-1}$, is assumed to play a negligible role here.
In any case,
inside this ``bag" it is quite possible that the electroweak symmetry
may
be
restored, in  which case such a
crust would not be present.)  In a low lying s-wave state, the
annihilation time
is very short.  However, in an excited state, involving, for example,
high
orbital angular momentum, this may not be the case, since the wave
function
at the origin will be highly suppressed. We provide here one approximate
estimate for the annihilation time based on the observation that the
Coulomb
capture distance $ a_c \approx 1/4\alpha E$ is 8 times smaller than the
``bag"
size, for a monopole whose ``bag" energy  is inferred from equation
(\ref{kappa})
with $L=2r$.  It is reasonable to suppose that annihilation might
proceed
via collapse into a tightly bound Coulomb state.
Thus, for the sake of argument one might roughly
estimate
a lower limit on the annihilation time by utilizing the Coulomb capture
cross
section\cite{Preskill}
inside
the ``bag".  This capture time is  $\tau \approx (4 e
/3\pi T) (m_M/T)^{11/10}$, and is slightly longer
than
the lifetime of the universe at temperature $T \approx 300 GeV$, for
$m_M=10^{17}
GeV$.      Again, this suggests that the time during which the
$U(1)_{em}$
breaking
phase endures must be long compared to the lifetime of the universe
when this phase begins\footnote{If one imagines that because of the
monopole outer crust, emission of scalars is possible, the
capture cross section may be increased\cite{Preskill} to $ \approx
(T_c)^{-2}$.
This  would
decrease the capture time by a significant amount ($\approx 10^{6}$).
However,
once again, this requires that the scalars are light, otherwise phase
space
suppression might be important.}.  If capture into a Coulomb state has
not
occured  by the time the $U(1)_{em}$ breaking phase is over,
previously confined monopole pairs separated by more than the
Coulomb capture distance will no longer be bound.
The annihilation rate  for these previously confined pairs
compared to the expansion rate will remain less than order
unity, so that monopoles will again freeze out

These considerations suggest that monopole-antimonopole
annihilation by
flux tube formation at the electroweak scale is far from guaranteed. In
particular, monopole confinement triggered by monopole induced
magnetic fields seems unworkable. More generally, in any confining
scenario,
dissipation of the initially large flux tube energies requires times
which are
generally long compared to the horizon time at the epoch of electroweak
symmetry breaking. This places strong constraints on the minimum range
of
temperatures over which a confining phase for monopoles must exist.

\vfill\eject

\noindent \medskip\centerline{\bf Figure Captions}
\vskip 0.15 truein
Fig. 1. Phase diagram for $W$ condensation as a function of
external
magnetic field and temperature assuming a second order electroweak phase
transtion.

\bibliographystyle{unsrt}

\begin{thebibliography}{99}

\bibitem{'tHooft}G. 't Hooft, {\em Nucl. Phys.} {\bf B79}, 276 (1974);
A. Polyakov, {\em JETP Lett.} {\bf 20}, 194 (1974).

\bibitem{Kibble}T.W.B. Kibble, {\em J. Phys.} {\bf A9}, 1387 (1976).

\bibitem{Preskill}J. Preskill, {\em Phys. Rev. Lett.}
{\bf 43}, 1365 (1979).

\bibitem{Zeldovich}Ya.A. Zeldovich and M.Y. Khlopov, {\em Phys. Lett.}
{\bf 79B}, 239 (1978).

\bibitem{parker}E.N. Parker, {\em Ap. J.} {\bf 160}, 383 (1970);
Y. Raphaeli and M.S. Turner, {\em Phys. Lett.} {\bf B121} 115 (1983);
M.E. Huber et. al., {\em Phys. Rev. Lett.} {\bf 64} 835 (1990);
S. Bermon et. al, {\em Phys. Rev. Lett.} {\bf 64} 839 (1990).

\bibitem{guth}A. Guth, {\em Phys. Rev.} {\bf D23}, 347 (1981).

\bibitem{LangPi}P. Langacker and S.-Y. Pi, {\em Phys. Rev. Lett.}
{\bf 45}, 1 (1980).

\bibitem{Lazarides}G. Lazarides and Q. Shafi, {\em Phys. Lett.}
{\bf B 94}, 149 (1980).

\bibitem{Owen}E.I. Guendelman and D.A. Owen, {\em Phys. Lett.}
{\bf B 235}, 313 (1990).

\bibitem{Sher}V.V. Dixit and M. Sher, {\em Phys. Rev. Lett.} {\bf 68},
560
(1992); T.H. Farris et. al., {\em Phys. Rev. Lett.} {\bf 68}, 564
(1992).

\bibitem{Skalozub}V.V. Skalozub, {\em Sov. J. Nucl. Phys.}
{\bf 23}, 113 (1978);
N.K. Nielsen and P. Olesen, {\em Nucl. Phys}
{\bf B144}, 376 (1978);
J. Ambj{\o}rn, R.J. Hughes, and N.K. Nielsen, {\em Ann. Phys. (N.Y.)}
{\bf 150}, 92 (1983).

\bibitem{AO1990}J. Ambj{\o}rn and P. Olesen, {\em Nucl. Phys.}
{\bf B 330}, 193 (1990).

\bibitem{AO1989a}J. Ambj{\o}rn and P. Olesen, {\em Nucl. Phys.}
{\bf B 315}, 606 (1989).

\bibitem{AO1991}J. Ambj{\o}rn and P. Olesen, {\em Phys. Lett.}
{\bf B 257}, 201 (1991); J. Ambj{\o}rn and P. Olesen, {\em Phys. Lett.}
{\bf B 218}, 67 (1989); J. Ambj{\o}rn and P. Olesen, {\em Phys. Lett.}
{\bf B 214}, 565 (1991).

\bibitem{Bogo}E.B. Bogomol'nyi, {\em Sov. J. Nucl. Phys.}
{\bf 24}, 449 (1977).

\bibitem{Vilenkin}A. Vilenkin, {\em Nucl. Phys.} {\bf B196}, 240 (1982).
\bibitem{alford} M. Alford and F. Wilczek, {\em Phys. Rev. Lett.} {\bf
62}, 1071
(1989)
\bibitem{sj}  R. Holman, T. Kibble, S.-J. Rey, preprint, YCTP-P06-92
(Yale),
NSF-ITP-09 (Santa Barbara) to appear in {\em Phys. Rev. Lett.}
\end{thebibliography}

 \end{document}